\documentclass[twocolumn,showpacs,preprintnumbers]{revtex4}%
\usepackage{amsfonts}
\usepackage{amssymb}
\usepackage{amsmath}
\usepackage{graphicx}
\usepackage{dcolumn}
\usepackage{bm}%

\begin{document}

\title{Correlation enhanced phase sensitive Raman scattering in atomic vapors%
}
\author{Chun-Hua Yuan$^{1,2}$, L. Q. Chen$^{1,2}$, Z. Y. Ou$^{1,2,3}$, and Weiping
Zhang$^{1,2\dag}$}
\affiliation{$^{1}$Quantum Institute for Light and Atoms, East China Normal University,
Shanghai 200062, P. R. China}
\affiliation{$^{2}$Department of Physics, State Key Laboratory of Precision Spectroscopy,
East China Normal University, Shanghai 200062, P. R. China}
\affiliation{$^{3}$Department of Physics, Indiana University-Purdue University
Indianapolis, 402 North Blackford Street, Indianapolis, Indiana 46202, USA}
\date{\today }

\begin{abstract}
We theoretically propose a method to enhance Raman scattering by injecting
a seeded light field which is correlated with the initially prepared atomic
spin wave. Such a light-atom correlation leads to an interference in the Raman
scattering. The interference is sensitive to the relative phase between the
seeded light field and initially prepared atomic spin wave. For constructive
interference, the Raman scattering is greatly enhanced. Such an enhanced Raman
scattering may find applications in quantum information, nonlinear optics and
optical metrology due to its simplicity.

\end{abstract}

\maketitle

\section{Introduction}

The cooperative spontaneous emission of radiation (superradiance) from an
ensemble was first introduced by Dicke in 1954 \cite{Dicke}, where an atomic
ensemble exhibited enhanced coupling to a single electromagnetic mode.
Superradiance was initially suggested for sample dimensions $R$ much smaller
than the wavelength $\lambda$ of the resonant transition \cite{Dicke}.
However, the case in the opposite limit $R\gg\lambda$ has also attracted
extensive research
\cite{Smithey,Jain,Black,Simon07,Simon,Scully09,Sete,Chen09,Chen10,Agarwal11}
because in quantum optics the sample is usually large compared to $\lambda$.
In such a large-size sample, the superradiance is difficult to happen
according to the standard argument that the dipole-dipole interaction between
atoms is too weak to build a macroscopic dipole moment. To have enhancement in
the limit of large-size sample with $R\gg\lambda$, the quantum coherence and
interference must enter to play a role. Typical example is to employ the
quantum coherence to enhance some nonlinear optical processes
\cite{Lukin,Boyd}. Recently, our group observed an enhanced Raman scattering
effect by using an atomic spin wave, a coherence prepared between atomic
ground-state sublevels \cite{Chen09,Chen10}. The subsequent theoretical
analysis modeled the experiment where an early spontaneous Raman scattering
(SRS) generated the atomic spin wave which causes enhancement of the Raman
light fields in the second Raman scattering \cite{Yuan10}. The flipped atoms
act as seeds to the second Raman process which amplifies the light fields all
the way through the atomic ensembles.

In this paper, based on our previous work \cite{Chen09,Chen10,Yuan10}, we
propose a scheme to enhance the Raman scattering, termed as
correlation-enhanced Raman scattering (CERS). In the scheme, a pump field
leads to spontaneous emission of the Stokes field, accompanying with the
generation of atomic spin waves. Then this Stokes field used as a seeded
signal, with the pump field together, is subsequently input into atomic
ensemble to generate a second Stokes field. A CERS occurs due to the
correlation of the seeding Stokes field and atomic spin waves
\cite{Smithey,Raymer04,Wasilewski,Ji07,Bian}. Such a light-atom correlation is
a new mechanism to enhance Raman scattering, which is different from the idea
of the so-called super Raman scattering initiated by the atom-atom entangled
state \cite{Agarwal11}.

Our article is organized as follows. In Sec.~\ref{RamanA}, the general model
involving spatial propagation and light-atom coupling in Raman scattering is
reviewed. The correlation of the light field and the atomic spin wave is
derived from the light-atom coupling equations. Based on the model, the CERS
is studied in detail in Sec.~\ref{RamanB}. In Sec.~\ref{analysis}, we
numerically calculate the intensity of CERS based on the correlation of the
seeded field and the atomic spin wave. Finally, we conclude with a summary of
our results.

\section{Light-atom Correlation in Raman Scattering}

\label{RamanA}

\begin{figure}[ptb]
\centerline{\includegraphics[scale=0.65,angle=0]{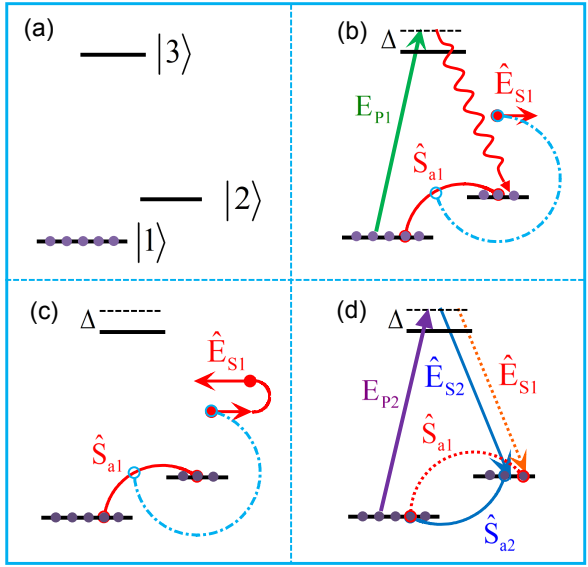}}\caption{(Color
online) (a) Initially prepared ground state for SRS. (b) SRS process: a Stokes
field and its correlated atomic spin wave are generated. The light-atom
correlation is denoted by the circular arc (dotted-dashed blue line). (c) The
light field and its correlated atomic spin wave, generated in the SRS process, are
used as initial seeding for CERS. (d) CERS process: an enhanced Stokes field
$\hat{E}_{S{2}}$ is generated by the pump $E_{P{2}}$ with the initial seeding
state (c). $E_{P{1,2}}$: pump fields; $\hat{E}_{S{1,2}}$ and $\hat{S}_{a{1,2}%
}$: the generated Stokes fields and spin waves, respectively; $\Delta$: the
detuning.}%
\label{fig1}%
\end{figure}

In this section, we give a brief review of the theoretical model of Raman
scattering. The Raman scattering process in atomic system is described through
a three-level model [see Fig.\ref{fig1}(a)] coupled to a pump field and a
Stokes field. For convenience, assume the pump field corresponding to a
focused beam and the Fresnel number $\mathfrak{F}=A/\lambda_{S}L$ ($A$
cross-sectional area, $L$ cell length) is of the order of unity, then only a
single transverse spatial mode contributes strongly to emission along the
direction of propagation the pump field. Therefore, a simplified one-dimensional model can be
enough to describe the Raman scattering. In the case of large light detuning,
the atomic excited state can adiabatically be eliminated and one obtains the
light-atom coupling equations governing the propagation of the quantized
Stokes field and the atomic ground-state spin excitations determined by the
spin wave \cite{Yuan10,Raymer81}
\begin{align}
&  (\partial_{t}+c\partial_{z})\hat{\mathcal{E}}_{S}(z,t)=i\chi\hat{S}%
_{a}^{\dag},\label{eq1}\\
\partial_{t}\hat{S}_{a}^{\dag}(z,t)  &  =-\Gamma_{S}\hat{S}_{a}^{\dag
}-i\mathcal{W}\chi^{\ast}\hat{\mathcal{E}}_{S}+\hat{F}_{S}^{\dagger
},\label{eq2}\\
\partial_{t}\mathcal{W}(z,t)  &  =-\gamma_{L}\mathcal{W}-\gamma_{L},
\label{eq3}%
\end{align}
where $\hat{S}_{a}=\sqrt{N}\tilde{\sigma}_{12}e^{-i\Delta kz}$ the spin wave
operator, $N$ the number of atoms, $\Delta k=k_{P}-k_{S}$, $\tilde{\sigma
}_{\mu\nu}(z,t)=1/N_{z}\sum_{j=1}^{N_{z}}\hat{\sigma}_{\mu\nu}^{j}$ the
collective atomic operators, where $\hat{\sigma}_{\mu\nu}^{j}=|\mu\rangle
_{jj}\langle\nu|$ ($\mu,\nu=1,2$) is the transition operator of the $j$th atom
between states $|\mu\rangle$ and $|\nu\rangle$ and a small and macroscopic
volume containing $N_{z}$ $(N_{z}\gg1)$ atoms around position $z$, and the
commutation relation is $[\tilde{\sigma}_{\mu\nu}(z,t),\tilde{\sigma}%
_{\alpha\beta}(z^{\prime},t)]=L/N\delta(z-z^{\prime})(\delta_{\nu\alpha}%
\tilde{\sigma}_{\mu\beta}(z,t)-\delta_{\mu\beta}\tilde{\sigma}_{\alpha\nu
}(z,t))$, $L$ the length of the atomic medium. $\hat{\mathcal{E}}_{S}(z,t)$ is
a slowly varying envelope operator of the Stokes field $\hat{E}_{S}%
^{(+)}=\sqrt{\hbar\omega/2\varepsilon_{0}AL}\hat{\mathcal{E}}_{S}%
(z,t)e^{i(\mathbf{k}_{S}\cdot\mathbf{z}-\omega_{S}t)}$, and $A$ is
cross-section area, and its commutation relation $[\hat{\mathcal{E}}%
_{S}(z,t),\hat{\mathcal{E}}_{S}(z,t^{\prime})]=L/c\delta(t-t^{\prime}%
)$.$\ \chi(z,t)=g\sqrt{N}\Omega_{P}(z,t)/\Delta$ the coupling coefficient
between spin excitations and stokes field, $\Omega_{P}(z,t)$\ the Rabi
frequency of pump field, $g$ the atom-field coupling constant, $\Gamma
_{S}=\gamma_{S}-i\delta_{L}$, $\gamma_{S}=\gamma_{S_{0}}+\gamma_{L}$, and
$\gamma_{L}=\gamma|\Omega_{P}|^{2}/\Delta^{2}$ is the optical pumping rate,
and $\delta_{L}=|\Omega_{P}|^{2}/\Delta$ is the ac Stark shift, and
$\gamma_{S_{0}}$ the coherence ($\tilde{\sigma}_{12}$) decay rate, $\gamma$
the decay rates of the excited state $|3\rangle$ to states $|1\rangle$ and
$|2\rangle$ (assuming $\gamma_{31}=\gamma_{32}=\gamma$). $\mathcal{W}%
(z,t)=\tilde{\sigma}_{11}(z,t)-\tilde{\sigma}_{22}(z,t)$ describes the
population difference between energy levels $|1\rangle$ and $|2\rangle$. In
general, $\mathcal{W}(z,t)$ is related to the collective atomic excitation
number and the strength of atomic coherence. Here we consider weak
excitations, and $\mathcal{W}(z,t)$ is approximately determined by Eq.
(\ref{eq3}). The Langevin noise operator $\hat{F}_{S}^{\dagger}$ has the
correlation $\langle\hat{F}_{S}(z,t)\hat{F}_{S}^{\dagger}(z^{\prime}%
,t^{\prime})\rangle=2\gamma_{S}L\delta(z-z^{\prime})\delta(t-t^{\prime})$.

For convenience in analysis, neglecting the depletion of the pump field by
making $\Omega_{P1,2}(z,t)\equiv\Omega_{P1,2}(t)=\Omega_{P1,2}\theta(t)$ with
$\Omega_{P1,2}$ being constant and $\theta(t)$ the step function. Similarly,
$\chi(z,t)\equiv\chi(t)=g\sqrt{N}\Omega_{P}(t)/\Delta$. Using the moving
coordinates $t^{\prime}=t-z/c$, $z^{\prime}=z$, the solutions of
Eqs.~(\ref{eq1})-(\ref{eq3}) are given for the Raman scattering
\cite{Yuan10,Raymer81},
\begin{align}
&  \hat{\mathcal{E}}_{S}(z^{\prime},t^{\prime})=\mathcal{U}_{S}(z^{\prime
},t^{\prime})\{\hat{\mathcal{E}}_{S}(0,t^{\prime\prime})\}+\mathcal{V}%
_{S}(z^{\prime},t^{\prime})\{\hat{S}_{a}^{\dagger}(z^{\prime\prime
},0)\}\nonumber\\
&  \text{ \ \ \ \ \ \ \ \ \ \ \ \ \ \ \ \ \ \ \ \ \ \ }+\mathcal{F}%
_{S}(z^{\prime},t^{\prime})\{\hat{F}_{S}^{\dagger}(z^{\prime\prime}%
,t^{\prime\prime})\},\label{solu1}\\
&  \hat{S}_{a}^{\dag}(z^{\prime},t^{\prime})=\mathcal{U}_{a}(z^{\prime
},t^{\prime})\{\hat{S}_{a}^{\dagger}(z^{\prime\prime},0)\}+\mathcal{V}%
_{a}(z^{\prime},t^{\prime})\{\hat{\mathcal{E}}_{S}(0,t^{\prime\prime
})\}\nonumber\\
&  \text{ \ \ \ \ \ \ \ \ \ \ \ \ \ \ \ \ \ \ \ \ \ \ }+\mathcal{F}%
_{a}(z^{\prime},t^{\prime})\{\hat{F}_{S}^{\dagger}(z^{\prime\prime}%
,t^{\prime\prime})\},\label{solu2}\\
&  \mathcal{W}(t^{\prime})=\mathcal{W}(0)e^{-\Gamma_{L}(t^{\prime})}-\int
_{0}^{t^{\prime}}\gamma_{L}(t^{\prime\prime})e^{-[\Gamma_{L}(t^{\prime
})-\Gamma_{L}(t^{\prime\prime})]}dt^{\prime\prime},\nonumber\\
&  \label{solu3}%
\end{align}
where%
\begin{align}
&  \mathcal{U}_{S}(z^{\prime},t^{\prime})\{...\}=\int_{0}^{t^{\prime}%
}dt^{\prime\prime}[2\delta(t^{\prime}-t^{\prime\prime})+\chi(t^{\prime}%
)\chi^{\ast}(t^{\prime\prime})\nonumber\\
&  ~~\text{\ \ \ \ \ \ \ \ \ }\times e^{-[\Gamma(t^{\prime})-\Gamma
(t^{\prime\prime})]}\frac{\mathcal{W}(t^{\prime\prime})}{c}G_{e}(z^{\prime
},0,t^{\prime},t^{\prime\prime})]\{...\},\nonumber\\
&  \mathcal{V}_{S}(z^{\prime},t^{\prime})\{...\}=i\int_{0}^{z^{\prime}%
}dz^{\prime\prime}\frac{\chi(t^{\prime})}{c}e^{-\Gamma(t^{\prime})}%
H(z^{\prime},z^{\prime\prime},t^{\prime},0)\{...\},\nonumber
\end{align}%
\begin{align}
&  \mathcal{U}_{a}(z^{\prime},t^{\prime})\{...\}=\int_{0}^{z^{\prime}%
}dz^{\prime\prime}e^{-\Gamma(t^{\prime})}[2\delta(z^{\prime}-z^{\prime\prime
})\nonumber\\
&  \text{ \ \ \ \ \ \ \ \ \ \ }\times G_{S}(z^{\prime},z^{\prime\prime
},t^{\prime},0)]\{...\},\nonumber\\
&  \mathcal{V}_{a}(z^{\prime},t^{\prime})\{...\}=-i\int_{0}^{t^{\prime}%
}dt^{\prime\prime}\mathcal{W}(t^{\prime\prime})\chi^{\ast}(t^{\prime\prime
})e^{-[\Gamma(t^{\prime})-\Gamma(t^{\prime\prime})]}\nonumber\\
&  \text{ \ \ \ \ \ \ \ \ \ \ }\times H(z^{\prime},0,t^{\prime},t^{\prime
\prime})\{...\},\nonumber\\
&  \mathcal{F}_{S}(z^{\prime},t^{\prime})\{...\}=i\int_{0}^{t^{\prime}%
}dt^{\prime\prime}\int_{0}^{z^{\prime}}dz^{\prime\prime}\frac{\chi(t^{\prime
})}{c}e^{-[\Gamma(t^{\prime})-\Gamma(t^{\prime\prime})]}\nonumber\\
&  \text{ \ \ \ \ \ \ \ \ \ \ }\times H(z^{\prime},z^{\prime\prime},t^{\prime
},t^{\prime\prime})\{...\},\nonumber\\
&  \mathcal{F}_{a}(z^{\prime},t^{\prime})\{...\}=\int_{0}^{t^{\prime}%
}dt^{\prime\prime}\int_{0}^{z^{\prime}}dz^{\prime\prime}e^{-[\Gamma(t^{\prime
})-\Gamma(t^{\prime\prime})]}\nonumber\\
&  \text{ \ \ \ \ \ \ \ \ \ \ }\times\lbrack2\delta(z^{\prime}-z^{\prime
\prime})+G_{S}(z^{\prime},z^{\prime\prime},t^{\prime},t^{\prime\prime
})]\{...\},
\end{align}
in which
\begin{align}
&  H(z^{\prime},z^{\prime\prime},t^{\prime},t^{\prime\prime})=I_{0}%
(2\sqrt{[q(t^{\prime})-q(t^{\prime\prime})]\frac{z^{\prime}-z^{\prime\prime}%
}{c}}),\nonumber\\
&  G_{e}(z^{\prime},z^{\prime\prime},t^{\prime},t^{\prime\prime}%
)=\frac{c(z^{\prime}-z^{\prime\prime})}{q(t^{\prime})-q(t^{\prime\prime}%
)}G_{S}(z^{\prime},z^{\prime\prime},t^{\prime},t^{\prime\prime}),\nonumber\\
&  G_{S}(z^{\prime},z^{\prime\prime},t^{\prime},t^{\prime\prime})=\sqrt
{\frac{q(t^{\prime})-q(t^{\prime\prime})}{c(z^{\prime}-z^{\prime\prime})}%
}\nonumber\\
&  \times I_{1}(2\sqrt{[q(t^{\prime})-q(t^{\prime\prime})]\frac{z^{\prime
}-z^{\prime\prime}}{c}}).
\end{align}
Here $q(t^{\prime})=\int_{0}^{t^{\prime}}\mathcal{W}(t^{\prime\prime}%
)\chi(t^{\prime\prime})^{2}dt^{\prime\prime}$, $\Gamma(t^{\prime})=\int
_{0}^{t^{\prime}}\Gamma_{S}(t^{\prime\prime})dt^{\prime\prime}$, $\Gamma
_{L}(t^{\prime})=\int_{0}^{t^{\prime}}\gamma_{L}(t^{\prime\prime}%
)dt^{\prime\prime}$, and $I_{n}(x)$\ is the modified Bessel function of the
first kind of order $n$.

The integral solutions presented in Eqs.~(\ref{solu1})-(\ref{solu2}) indicate
that the Stokes field $\hat{\mathcal{E}}_{S}(z^{\prime},t^{\prime})$ contains
three parts of contributions, the first one from the input field plus the
scattering field, the second from the initial spin excitations, and the third
from the atomic fluctuations. Similarly, the generated spin wave $\hat{S}%
_{a}^{\dagger}(z^{\prime},t^{\prime})$ contains the contributions from the
initial spin excitations, the input field, and the atomic fluctuations.
Evidently a light-atom correlation of the Stokes field and the generated spin
wave, $\hat{\langle\mathcal{E}}_{S}(z,t^{\prime})\hat{S}_{a}(z^{\prime
},t^{\prime})\rangle\neq0,$ is built through the Raman coupling as seen
in\textit{ }Eqs. (\ref{solu1})-(\ref{solu2}).

\section{Correlation-enhanced Raman Scattering}

\label{RamanB}

In the section, we study the role of the correlation of the light field and
atomic spin wave in enhancing Raman scattering. The basic idea is as follows.
First we employ a SRS process to generate a light field and an atomic spin
wave which are correlated with each other. The SRS process is illustrated in Figs.
\ref{fig1}(a) and (b), where an atomic ensemble, initially prepared in the ground
state, is pumped by an off-resonance pump field $E_{P1}$ and spontaneously
emitted a Stokes field $\hat{E}_{S1}$. As a result, a coherent spin excitation
$\hat{S}_{a1}$ is built in the atomic ensemble. Based on the analysis given in
the above section, the emitted Stokes field $\hat{E}_{S1}$ is correlated with
the excited spin wave $\hat{S}_{a1}.$ Next, we consider a new Raman scattering
process, termed as CERS, where the Stokes field $\hat{E}_{S1}$, with its
correlated atomic spin wave $\hat{S}_{a1}$, is employed as initial seeding as
shown in Fig. \ref{fig1}(c). Again an off-resonance pump field $E_{P2}$ is
injected to generate a new Stokes emission $\hat{E}_{S2}$, as illustrated in
Fig. \ref{fig1}(d).

\begin{figure}[ptb]
\centerline{\includegraphics[scale=0.75,angle=0]{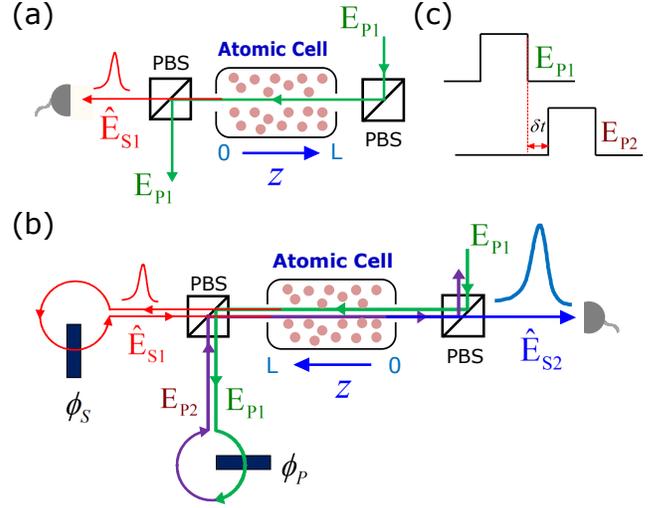}}\caption{(Color
online) Schematic diagram of experimental set up for (a) SRS: the pump field
$E_{P{1}}$ generates the Stokes field $\hat{E}_{S{1}}$ and the atomic spin
wave $\hat{S}_{a1}$ through spontaneous emission, and (b) CERS: after a delay
$\delta t$, the leaked pump $E_{P{1}}$, acting as $E_{P{2}}$, is again
injected into the vapor cell with the SRS-generated Stokes field $\hat
{E}_{S{1}}$. The SRS-generated correlation of the light $\hat{E}_{S{1}}$ and
the spin wave $\hat{S}_{a1}$ enhances the intensity of the second Raman Stokes
field $\hat{E}_{S{2}}$. PBS: polarization beam splitter; $\phi_{P}$, $\phi
_{S}$: phase shifts. (c) Time sequence for the pump pulses. $\delta t$: delay
time.}%
\label{fig2}%
\end{figure}

The experiment to study the CERS can simply be setup as in Fig.~\ref{fig2}(b).
An atomic cell, for example, rubidium vapor cell is used to provide the atomic
ensemble. The generated Stokes field $\hat{E}_{S1}$ and the leaked pump
$E_{P1}$ in the SRS, experiencing phase shifts $\phi_{S}$ and $\phi_{P}$
respectively, are injected back into the atomic cell. The light-atom
correlation are used as the initial seeding and the the leaked pump acting as
$E_{P2}$ for the CERS. The output intensity of the Stokes field $\hat{E}_{S2}$
is detected for analysis.

Using the relations given in Eqs. (\ref{solu1})-(\ref{solu3}), we can compare
the differences between the SRS process and CERS process by working out their
output intensity in detail. In the case of the SRS, the Stokes field is generated
from the vacuum and the atomic ground state without light-atom correlation.
Hence the initial conditions for the SRS are$\ $
\begin{align}
\langle\hat{\mathcal{E}}_{S1}^{\dag}(0,t^{\prime})\hat{\mathcal{E}}%
_{S1}(0,t^{\prime})\rangle &  =\langle\hat{S}_{a1}^{\dag}(z^{\prime},0)\hat
{S}_{a1}(z^{\prime},0)\rangle=0,\\
\langle\hat{\mathcal{E}}_{S1}(0,t^{\prime})\hat{S}_{a1}(z^{\prime},0)\rangle
&  =0.
\end{align}
where the subscript $1$ denotes the SRS process. The detected intensity of the
SRS output Stokes field at the end of the atomic cell $z^{\prime}=L$ is given
by
\begin{align}
I_{\text{out}}^{\text{SRS}}  &  =\langle\hat{E}_{S1}^{(-)}\hat{E}_{S1}%
^{(+)}\rangle\varpropto\langle\hat{\mathcal{E}}_{S1}^{\dag}(L,t^{\prime}%
)\hat{\mathcal{E}}_{S1}(L,t^{\prime})\rangle\nonumber\\
&  =L\left\vert \mathcal{V}_{S1}(L,t^{\prime})\right\vert ^{2}+\left\vert
\mathcal{F}_{S1}(L,t^{\prime})\right\vert ^{2}\nonumber\\
&  \times\langle\hat{F}_{S1}(z^{\prime\prime},t^{\prime\prime})\hat{F}%
_{S1}^{\dagger}(z^{\prime\prime},t^{\prime\prime})\rangle\nonumber\\
&  \equiv I_{\text{spon}_{\text{1}}}. \label{SRS}%
\end{align}
The result reflects the fact that the Stokes field is generated from the
spontaneous emission.

However, in the case of the CERS, one has a completely different initial
condition as follows:%
\begin{equation}
\hat{\mathcal{E}}_{S2}(0,t^{\prime})=\hat{\mathcal{E}}_{S1}(L,t^{\prime
}),\text{ }\hat{S}_{a2}^{\dagger}(z^{\prime},0)=\hat{S}_{a1}^{\dagger
}(L-z^{\prime},T), \label{initial-stokes}%
\end{equation}
where the subscript $2$ denotes the CERS process, $L-z^{\prime}$\ indicates
the opposite propagation of the two pump fields, and $T$ is the pulse duration
of the pump $E_{P1}$. Evidently from Eq. (\ref{SRS})\textit{,} the light-atom
correlation will play an important role in the CERS due to the relations
\begin{align}
\langle\hat{\mathcal{E}}_{S2}^{\dag}(0,t^{\prime})\hat{\mathcal{E}}%
_{S2}(0,t^{\prime})\rangle &  \neq0,\text{ }\langle\hat{S}_{a2}^{\dag
}(z^{\prime},0)\hat{S}_{a2}(z^{\prime},0)\rangle\neq0,\text{ }\\
\langle\hat{\mathcal{E}}_{S2}(0,t^{\prime})\hat{S}_{a2}(z^{\prime},0)\rangle
&  \neq0.
\end{align}
In general, the spin wave $\hat{S}_{a1}(z^{\prime},T)$ decays due to
collisional dephasing by a factor $e^{-\Gamma_{S}\delta t}$ with $\delta t$
being the delay time. Here for convenience, we neglect the slow decay of the
spin wave by assuming a short delay time ($\Gamma_{S}\delta t\ll1$).

\begin{figure}[ptb]
\centerline{\includegraphics[scale=0.5,angle=0]{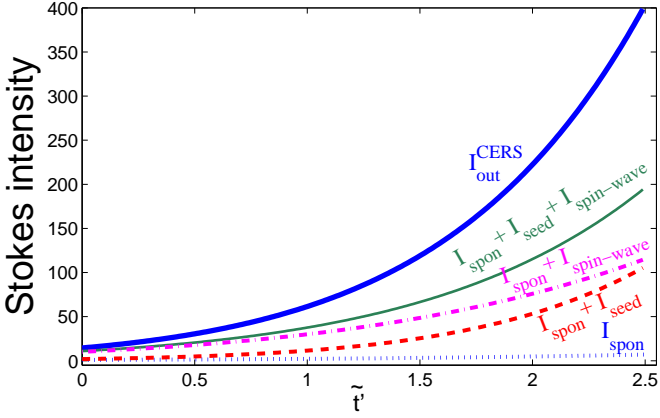}}\caption{(Color
online) The intensity of Stokes field $\hat{\mathcal{E}}_{S{2}}$ versus the
dimensionless time $\tilde{t}^{\prime}$: SRS case (dotted line), seeded-light
only case (dashed line), spin-wave only case (dotted-dashed line), light-atom
uncorrelated case (thin solid line), and light-atom correlated case (thick
solid line). The parameters are as follows: $\Delta\phi+\Delta\phi
_{\text{offset}}=0$, $\mathcal{W}_{2}(0)=0.99$, $\Delta=1.2$ GHz,
$|\Omega_{P{1}}|=|\Omega_{P{2}}|=2.5\times10^{8}$ Hz, $Ng_{2}^{2}%
/c=1\times10^{12}/(m\cdot s)$ , $\gamma_{s}=10 $ KHz, and $\gamma_{1}%
=2\pi\times5.746$ MHz. }%
\label{fig3}%
\end{figure}

The detected intensity of the CERS output Stokes field can be worked out
\begin{align}
I_{\text{out}}^{\text{CERS}}  &  =\langle\hat{E}_{S2}^{(-)}\hat{E}_{S2}%
^{(+)}\rangle\varpropto\langle\hat{\mathcal{E}}_{S2}^{\dag}(L,t^{\prime}%
)\hat{\mathcal{E}}_{S2}(L,t^{\prime})\rangle\nonumber\\
&  =I_{\text{spon}}+I_{\text{seed}}+I_{\text{spin-wave}}+I_{\text{light-atom}%
}, \label{CERS}%
\end{align}
where
\begin{align}
&  I_{\text{spon}}=L\left\vert \mathcal{V}_{S2}(L,t^{\prime})\right\vert
^{2}+\left\vert \mathcal{F}_{S2}(L,t^{\prime})\right\vert ^{2}\nonumber\\
&  ~~~~~~\times\langle\hat{F}_{S2}(z^{\prime\prime},t^{\prime\prime})\hat
{F}_{S2}^{\dagger}(z^{\prime\prime},t^{\prime\prime})\rangle,\nonumber\\
&  I_{\text{seed}}=\left\vert \mathcal{U}_{S2}(L,t^{\prime})\right\vert
^{2}\langle\hat{\mathcal{E}}_{S1}^{\dag}(L,t^{\prime\prime})\hat{\mathcal{E}%
}_{S1}(L,t^{\prime\prime})\rangle,\nonumber\\
&  I_{\text{spin-wave}}=\left\vert \mathcal{V}_{S2}(L,t^{\prime})\right\vert
^{2}\nonumber\\
&  ~~~~~~\times\langle\hat{S}_{a1}^{\dagger}(L-z^{\prime\prime},T)\hat{S}%
_{a1}(L-z^{\prime\prime},T)\rangle,\nonumber\\
&  I_{\text{light-atom}}=e^{i\Delta\phi\text{ }}\mathcal{U}_{S2}^{\ast
}(L,t^{\prime})\mathcal{V}_{S2}(L,t^{\prime})\nonumber\\
&  ~~~~~~\times\langle\hat{\mathcal{E}}_{S1}^{\dag}(L,t^{\prime\prime})\hat
{S}_{a1}^{\dagger}(L-z^{\prime\prime},T)\rangle+\text{c.c.}%
\end{align}
The term $I_{\text{spon}}$ originates from the same mechanism as in the SRS.
The injected seeding field and the initial spin wave contribute the intensity
terms $I_{\text{seed}}$ and $I_{\text{spin-wave}}$. The light-atom correlation
leads to an interference term $I_{\text{light-atom}}$, which is sensitive to
the phase difference $\Delta\phi\equiv\phi_{P}-\phi_{S}$, experienced by the
pump and the Stokes field in CERS.

The CERS much depends on the initially-set conditions. A different setting
mechanism of initial conditions results in different enhancement. For the case
discussed here, this can be summarized as follows. When only seeded light
(spin wave) exists, the corresponding enhancement $I_{\text{seed}}$
($I_{\text{spin-wave}})$ occurs. Then the output intensity is
\begin{equation}
I_{\text{out}}=I_{\text{spon}}+I_{\text{seed (spin-wave)}},
\end{equation}
The spin wave (atomic coherence) induced enhancement mechanism was studied by
both theory and experiment \cite{Chen09,Yuan10,Chen10,Smithey}. When the
uncorrelated light field and the spin wave are used as the initial input
seeding, the enhancement is independently contributed by the light field and
the spin wave, respectively. As a result, the output intensity of the Stokes
field is a simple sum given as
\[
I_{\text{out}}=I_{\text{spon}}+I_{\text{seed}}+I_{\text{spin-wave}}.
\]
The light-atom correlation is a new mechanism which leads to the
phase-sensitive interference in the Raman scattering. It is well-known that
the constructive interference can greatly enhance the light intensity. In this
sense, the CERS occurs optimally for an appropriate phase difference
determined by $\Delta\phi+\Delta\phi_{\text{offset}}=2n\pi$ $(n=0,1...)$,
where $\Delta\phi_{\text{offset}}$ is an overall phase shift offset by the
seeded light, initial spin wave and the Raman scattering process itself. In
addition, we point out that the light-atom correlation also leads to a
maximally reduced Raman scattering when the phase difference satisfies the
condition $\Delta\phi+\Delta\phi_{\text{offset}}=(2n+1)\pi$ $(n=0,1...)$.

\bigskip

\section{Numerical Analysis}

\label{analysis}

In this section, we numerically calculate the intensities for different
mechanisms of enhancement. For convenience, we define the dimensionless time
$\tilde{t}^{\prime}=t^{\prime}\chi_{1}^{2}L/c=t^{\prime}\chi_{2}^{2}L/c$.
According to the different scattering mechanisms, the Stokes intensities of
different cases are plotted.

\begin{figure}[ptb]
\centerline{\includegraphics[scale=0.5,angle=0]{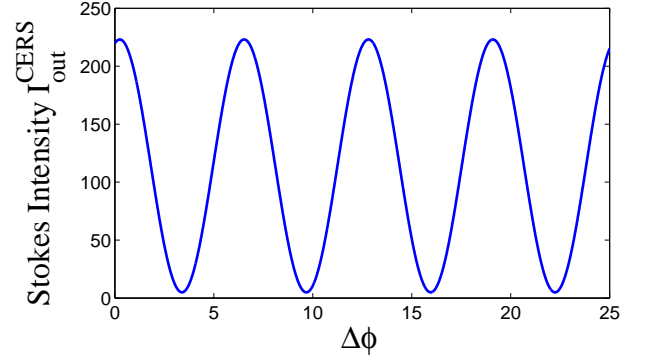}}\caption{(Color
online) The intensity of Stokes field $\hat{\mathcal{E}}_{S{2}}$ versus the
relative phase $\Delta\phi$. The other parameters are the same as those of
Fig.~\ref{fig3} except the dimensionless time at $\tilde{t}^{\prime}=2$.}%
\label{fig4}%
\end{figure}

In Fig.~\ref{fig3} the intensity determined by the SRS mechanism $I_{\text{spon}}%
$\ is plotted as the dotted line. The contributions from the seeded light
$I_{\text{seed}}$ and the spin wave $I_{\text{spin-wave}}$ are shown in the dashed
and the dotted-dashed lines, respectively. The enhanced total intensity without
the light-atom correlation $I_{\text{out}}=I_{\text{spon}}+I_{\text{seed}%
}+I_{\text{spin-wave}}$ is represented by the thin solid line. The optimal
CERS output intensity $I_{\text{out}}^{\text{CERS}}$ is evidently increased
due to the correlation-induced interference for the special phase difference
$\Delta\phi+\Delta\phi_{\text{offset}}=0$, chosen in the calculation, where
the result is displayed as a thick solid line.

Furthermore, we analyze the phase sensitivity of the CERS. In Fig.~\ref{fig4},
we plot the intensity $I_{\text{out}}^{\text{CERS}}$ as a function of the
phase difference $\Delta\phi$. This figure shows that the intensity is
modulated by the relative phase $\Delta\phi$. The modulation is a reflection
of the correlation-induced interference. The visibility of the interference
fringes is given by $V=[(I_{\text{out}}^{\text{CERS}})_{\max}-(I_{\text{out}%
}^{\text{CERS}})_{\min}]/[(I_{\text{out}}^{\text{CERS}})_{\max}+(I_{\text{out}%
}^{\text{CERS}})_{\min}]$. In the CERS, the high visibility can be achieved by
controlling the seeded light and initial spin wave to realize the case
$I_{\text{seed}}=I_{\text{spin-wave}}$. A direct application of the phase-sensitive Raman scattering proposed here is to realize a nonlinear
interferometer \cite{Yurke,Jing11}, which can benefit the improvement of phase
estimation for optical metrology, imaging, and information processing
\cite{Cave,Giovannetti,Dowling}.

\section{Conclusion}

In conclusion, we have theoretically studied the correlation-induced
phase-sensitive Raman scattering, which is based on the light-atom correlation
through the coupling of the light field with the atomic spin excitations in
the Raman scattering. We analyze the phase sensitivity of such a Raman
scattering process. An optimally enhanced Raman scattering occurs when the
accumulated phase difference determined by the pump, the seeded light, and the
Raman process is appropriate. Similarly, one can also have a reduced Raman
scattering by tuning the phase difference. Such a correlation-induced
phase-sensitive Raman scattering process may find applications in a diversity
of technological areas such as optical detection, metrology, imaging,
precision spectroscopy, and so on.

\begin{acknowledgements}
This work was supported by the National Basic Research Program of China (973
Program) under Grant No.~2011CB921604 and No.~11234003 (W.Z.), the National Natural Science
Foundation of China under Grant No.~11129402 (Z.Y.O.), No.~11004058, No.~11274118, and Supported by Innovation Program of Shanghai Municipal Education Commission 13zz036 (L.Q.C.), the National Natural Science Foundation of China under Grant No.~11004059 (C.H.Y.).
\newline
Email: $^{\dag }$wpzhang@phy.ecnu.edu.cn
\end{acknowledgements}

\end{document}